\documentclass{article}

\usepackage[preprint, nonatbib]{neurips_2021}

\usepackage[utf8]{inputenc} 
\usepackage[T1]{fontenc}    
\usepackage{hyperref}       
\usepackage{url}            
\usepackage{booktabs}       
\usepackage{amsfonts}       
\usepackage{nicefrac}       
\usepackage{microtype}      
\usepackage{xcolor}         

\usepackage{algorithm, algorithmic}
\usepackage{array}
\usepackage{multirow}
\usepackage{booktabs}
\usepackage[pdftex]{graphicx}
\usepackage{amssymb}   
\usepackage{amsmath}
\usepackage{amsthm}
\usepackage{mathtools}
\usepackage{enumitem}
\usepackage{subcaption}
\usepackage{pdfpages}

\title{Spatio-Temporal Momentum: Jointly Learning Time-Series and Cross-Sectional Strategies}

\author{
  Wee Ling Tan \\
  Department of Engineering Science\\
  Oxford-Man Institute of Quantitative Finance\\
  University of Oxford \\
  \texttt{weeling@robots.ox.ac.uk} \\
   \AND
   Stephen Roberts \\
  Department of Engineering Science\\
  Oxford-Man Institute of Quantitative Finance\\
  University of Oxford \\
  \texttt{sjrob@robots.ox.ac.uk} \\
  \AND
  Stefan Zohren \\
  Department of Engineering Science\\
  Oxford-Man Institute of Quantitative Finance\\
  University of Oxford \\
  \texttt{stefan.zohren@eng.ox.ac.uk}
}

\begin{document}

\maketitle

\begin{abstract}
We introduce Spatio-Temporal Momentum strategies, a class of models that unify both time-series and cross-sectional momentum strategies by trading assets based on their cross-sectional momentum features over time. While both time-series and cross-sectional momentum strategies are designed to systematically capture momentum risk premia, these strategies are regarded as distinct implementations and do not consider the concurrent relationship and predictability between temporal and cross-sectional momentum features of different assets. We model spatio-temporal momentum with neural networks of varying complexities and demonstrate that a simple neural network with only a single fully connected layer learns to simultaneously generate trading signals for all assets in a portfolio by incorporating both their time-series and cross-sectional momentum features. Backtesting on portfolios of 46 actively-traded US equities and 12 equity index futures contracts, we demonstrate that the model is able to retain its performance over benchmarks in the presence of high transaction costs of up to 5-10 basis points. In particular, we find that the model when coupled with least absolute shrinkage and turnover regularization results in the best performance over various transaction cost scenarios.
\end{abstract}

\section{Introduction}
\label{introduction}

Momentum strategies form a class of systematic strategies that rely on the premise of a persistence in the direction of asset returns over time \cite{moskowitz2012time, jegadeesh1993returns}. These strategies are constructed to exploit the continuation of the underlying trend by increasing or assuming long positions during uptrends, and decreasing or assuming short positions during downtrends. Momentum strategies are often accompanied by volatility targeting, allowing momentum strategies to leverage positions taken during periods of low volatility, and reduce exposures during periods of high volatility. Volatility targeting at both the asset and portfolio level has been shown to boost Sharpe ratios, reducing the probability of extreme tail returns and minimizing maximum drawdowns for risk asset portfolios \cite{harvey2018impact, kim2016time}.

Momentum strategies can be distinguished as belonging to a time-series or cross-sectional strategy. In terms of predictability, the former is primarily driven by the persistence in the trend of individual and market-level returns while the latter is usually deployed as a relative value strategy due to its perceived market-neutrality \cite{baz2015dissecting}. As such, current work has mainly regarded time-series and cross-sectional momentum as distinct implementations.

In time-series momentum strategies \cite{moskowitz2012time, garg2021momentum, lim2019enhancing, wood2022slow}, trading signals for individual assets are constructed based on the asset's own historical returns. In addition, the signals for individual assets in the portfolio are typically constructed independently of other assets. In the presence of a defined universe or portfolio of assets, this strategy does not take into consideration any form of mutual interactions and predictability between assets. We believe that momentum features from other assets represent sources of information as these assets collectively represent the market state at any point in time, and should be considered when constructing trading signals for a given asset.

On the other hand, cross-sectional momentum strategies \cite{jegadeesh1993returns, rouwenhorst1998international, pirrong2005momentum, poh2021building, poh2022enhancing}, require a momentum score to first be quantified for each individual asset in the portfolio, before computing a relative ranking of these scores in order to formulate positions for a select group of assets. In the first step of calculating a momentum score, cross-sectional momentum strategies ultimately consider only an asset's own historical returns, independent of returns from other assets. In principle, this is still identical to the way a trading signal is constructed in a time-series momentum strategy. Subsequently in the ranking step, a typical cross-sectional momentum strategy would allocate positions by assuming a maximum long position for assets ranked in the top decile, while taking a maximum short position for assets ranked in the bottom decile. We argue that this maximum long-short allocation does not accurately reflect the underlying trading signal for the selected assets. Moreover, this approach leaves a large majority of assets classified in intermediate deciles with no directional positions. As such, a cross-sectional strategy is unable to exploit any underlying trends for these intermediate assets.

In this work, we combine both types of strategies by considering a form of multi-asset momentum that simultaneously constructs trading signals based on momentum features from multiple assets over time. We model the spatio-temporal momentum strategy using a variety of neural networks trained in a data-driven manner, each representing different complexities. In predicting the trend and position size for any given asset, we consider the asset’s own momentum features as well as the momentum features from other assets in the portfolio. In contrast to a time-series momentum strategy, our model learns to incorporate information collectively from the universe of assets into signal construction. Our proposed strategy also directly generates trading signals for every asset as a multi-target output of the neural network, effectively bypassing the need to manually rank assets relative to one another as in the case of cross-sectional momentum. We directly optimize the models with the Sharpe ratio \cite{sharpe1998sharpe}, allowing the spatio-temporal momentum networks to learn from risk-adjusted performance \cite{lim2019enhancing}.

Our strategy closely resembles multitask learning \cite{caruana1997multitask, crawshaw2020multi}, a training framework that aims to train a learner on multiple tasks simultaneously. We observe that the prediction of momentum signals for different assets can be treated as multiple different, but closely related prediction tasks. In the context of spatio-temporal momentum, the model leverages on the usefulness of a shared feature representation between different assets, and is trained by optimizing over an aggregate Sharpe ratio computed from the multi-output model.

\section{Related Work}
\label{related_work}

Momentum strategies are typically classified as either time-series or cross-sectional. The work of \cite{moskowitz2012time} incorporates volatility scaling \cite{harvey2018impact} into a time-series momentum strategy and documents significant excess returns arising from trading based on the sign of an asset's own returns over the past 12 months, backtesting across 58 instruments and more than 25 years of data. Since then, other works have introduced more complex trend estimation techniques, such as using volatility normalised MACD signals \cite{baz2015dissecting} and blending slow and fast momentum strategies with various weights to reduce downside exposure \cite{garg2021momentum}. On the other hand, \cite{jegadeesh1993returns} show that a cross-sectional momentum strategy of buying winners and selling losers over a lookback horizon of 3 to 12 months led to significant excess returns for a portfolio of NYSE and AMEX stocks. The cross-sectional momentum effect has also been observed in international equity markets \cite{rouwenhorst1998international} and futures contracts \cite{pirrong2005momentum}.

Machine learning algorithms have been increasingly developed to perform predictions by extracting and modelling features from data. Deep learning has eliminated the need for manual feature engineering, allowing for hierarchical feature representations to be learnt directly from data \cite{lecun2015deep}. With recent advances in deep learning methods and open source libraries \cite{abadi2016tensorflow, paszke2019pytorch}, deep neural networks have been applied to model financial datasets such as high frequency limit order book data \cite{zhang2019deeplob} and to construct portfolios \cite{zhang2020deep}. Very often, neural networks that depend on recurrence and backpropagation through time, like recurrent neural networks (RNN) and long short-term memory networks (LSTM) \cite{hochreiter1997long}, are used to model temporal relationships commonly present in financial data.

Deep learning algorithms have also been applied to momentum strategies. The work of \cite{lim2019enhancing} introduces Deep Momentum Networks (DMNs), a series of neural network architectures directly optimised for portfolio Sharpe ratio, in place of standard regression and classification methods for time-series momentum. The models directly output positions for individual contracts, combining trend estimation and position sizing into a single function. However, we note that the single-output DMN model is trained on batches of input features belonging to different asset classes. This requires the model to learn to construct and output trading signals individually for all types of instruments, regardless of the possibility of negative transfer between asset classes. Our approach adopts a multitask learning approach that incorporates a multi-output target, permitting specialization for individual assets while retaining the advantage of having only a single model that leverages on a common shared representation between different assets.

Multitask learning \cite{caruana1997multitask, crawshaw2020multi, ruder2017overview} aims to train a model on multiple tasks simultaneously to improve generalization performance across tasks. It has has been successfully adopted in a range of applications, including computer vision \cite{zhang2014facial, misra2016cross, long2017learning} and natural language processing \cite{luong2015multi, dong2015multi, rei2017semi}. Training a deep learning-based multitask model involves backpropagation with training signals from multiple related tasks. This can be seen as a form of model regularization via shared representations, specifically by coupling a main task with auxiliary tasks to introduce an inductive bias to the model \cite{ruder2017overview}. We take the approach where all tasks (assets in the portfolio) are treated equally, unlike some multitask learning models that distinguish between a main task and one or more auxiliary tasks.

\section{Momentum Strategies}
\label{strategy_definition}
Following the definition of \cite{baltas2020demystifying, moskowitz2012time}, the overall returns of a time-series momentum (TSMOM) strategy that equally diversifies over $N_t$ assets at time $t$ is:

\begin{equation}
\label{eqn:tsmom}
r_{t,t+1}^{\text{TSMOM}} = \frac{1}{N_t} \sum_{i=1}^{N_t} X_t^{(i)}~\frac{\sigma_{\text{tgt}}}{\sigma_t^{(i)}}~r_{t,t+1}^{(i)}
\end{equation}

where $X_t^{(i)} \in [-1, 1]$ denotes the trading signal or position for asset $i$ at time $t$. Given the differences in volatility across individual assets, we scale the realized returns $r_{t,t+1}^{(i)}$ by their volatility to target equal risk assignments. We set the annualized volatility target $\sigma_{\text{tgt}}$ to be $15\%$ and estimate the ex-ante volatility $\sigma_t^{(i)}$ with a 60-day exponentially weighted moving standard deviation of daily returns.

Most momentum strategies are concerned with designing a proper trading signal $X_t^{(i)}$. We illustrate this with the following examples which we incorporate as benchmarks in our work:
\paragraph{Long Only} In the simplest case, a long only strategy takes a maximum long position $X_t^{(i)} = 1$  for each asset and performs a daily rebalancing according to changes in the ex-ante volatility estimate.

\paragraph{TSMOM} In the time-series momentum strategy of \textit{Moskowitz et al., 2012} \cite{moskowitz2012time}, the position taken for an asset is based on the sign of the asset's returns over the past 12 months: $X_t^{(i)} = \text{sgn}(r_{t-252, t}^{(i)})$

\paragraph{MACD} In \textit{Baz et al., 2015} \cite{baz2015dissecting}, volatility normalised moving average convergence divergence (MACD) indicators are used in place of the sign of returns as signals for trend estimation:
\begin{align}
    &\text{MACD}(i, t, S, L) = m(i, t, S) - m(i, t, L) \nonumber \\
    &\text{MACD}_{\text{norm}}(i, t, S, L) = \frac{\text{MACD}(i, t, S, L)}{\text{std}(p_{t-63 : t})} \nonumber \\
    &Y^{(i)}_t = \frac{\text{MACD}_{\text{norm}}(i, t, S, L)}{\text{std}(\text{MACD}_{\text{norm}}(i, t-252 : t, S, L))} \nonumber \\
    &X_t^{(i)} \coloneqq \tilde{Y}_t^{(i)} = \frac{1}{3} \sum_{k=1}^3 \phi(Y_t^{(i)} (S_k, L_k)) \label{eqn:macd}
\end{align}
where $\text{MACD}(i, t, S, L)$ is the MACD value of asset $i$ at time $t$ with a short time scale $S$ and long time scale $L$. Further, $m(i, t, j)$ is defined as the exponentially weighted moving average of asset $i$ prices at time $t$, with a time scale $j$ that corresponds to a half-life of $HL = \log(0.5) / \log(1 - \frac{1}{j})$. The MACD value is normalized by $\text{std}(p ^{(i)}_{t-63 : t})$, the 63-day rolling standard deviation of asset $i$ prices. Multiple intermediate MACD signals over different short and long time scales $S_{k} \in \{8, 16, 32\}$ and $L_{k} \in \{24, 48, 96\}$ are combined in an equally weighted sum to yield a position $X_t^{(i)}$ (or an aggregated MACD signal $\tilde{Y}_t^{(i)}$) where $\phi(y) = \frac{y\exp(\frac{-y^2}{4})}{0.89}$ is a response function as defined in \cite{baz2015dissecting}.

\paragraph{Deep Momentum Networks (DMN)} \textit{Lim et al., 2019} \cite{lim2019enhancing} use deep neural networks to directly generate trading signals for individual assets, combining trend estimation and position sizing into a single function as approximated by the learnt model $f$ using time-series momentum features $\mathbf{u}_t^{(i)}$:
\begin{align}
& X_t^{(i)} = f\left(\mathbf{u}_t^{(i)}; \boldsymbol{\theta} \right)
\end{align}
The model parameters $\boldsymbol{\theta}$ are learnt by optimizing over loss metrics that include volatility characteristics of returns. As a benchmark, we consider $f$ to be the Sharpe-optimized LSTM.

\paragraph{CSMOM} Following \textit{Jegadeesh \& Titman, 1993} \cite{jegadeesh1993returns}, we consider a cross-sectional momentum (CSMOM) strategy that scores and ranks an asset based on its returns computed over the past 12 months. The strategy utilizes a decile portfolio, taking a maximum long and short position for the top and bottom $10\%$ of ranked assets respectively.

\section{Spatio-Temporal Momentum}
\label{spatio_temporal_momentum}
\paragraph{Multitask Learning Premise} We observe that the prediction of individual momentum signals $X_t^{(i)}$ for different assets can be treated as multiple different, but closely related prediction tasks. In the context of an equally weighted portfolio, we take the approach where all assets (tasks) are treated equally in a multi-target setting. In addition, we follow the definition of multitask learning in \cite{caruana1997multitask}, where all tasks share the same inputs, unlike some definitions of multitask learning where different inputs are utilized for different tasks.

\subsection{Multitask Learning in Spatio-Temporal Momentum}

We model the simultaneous prediction of multiple trading signals $X_t^{(i)}$ with model $f$ as a multitask learning problem, with each task $i$ corresponding to predicting the trading signal for asset $i$. Given time $t$, our goal is to construct a multitask learning model $f$ that is able to perform $N^t$ tasks of predicting an aggregate trading signal $\mathbf{X}_t \in [-1, 1]^{N^t}$ over all assets.

In general, we have an input spatio-temporal tensor $\mathbf{u}_t \in \mathbb{R}^{N^t \times \tau \times d}$, where $N^t$ is the number of assets, $\tau$ is the temporal history, $d$ is the number of features, and $\mathbf{u}_t(i, j, k)$ represents the $k$-th feature of the $i$-th asset at time $t-j$. We want to learn a model $f$ parameterized by $\boldsymbol{\theta}$:
\begin{equation}
\label{eqn:input_to_signal}
\mathbf{X}_t = f(\mathbf{u}_t ; \boldsymbol{\theta})
\end{equation}
where
\begin{equation}
\mathbf{X}_t = 
\begin{bmatrix} 
X_t^{(1)} \\ 
X_t^{(2)} \\ 
\vdots \\ 
X_t^{(N^t)} \\
\end{bmatrix}
\end{equation}

In this framework, trading signals $X_t^{(i)}$ are directly computed using features from the universe of assets in the form of the spatio-temporal tensor $\mathbf{u}_t$. Specifically, this reduces to a time-series momentum strategy for $N^t = 1$. The realized returns of this framework still follows Equation (\ref{eqn:tsmom}).

\subsection{Deep Learning Architectures}
Given that the choice of model architecture is crucial in modelling the relationship between the trading signals and the spatio-temporal momentum features, we examine a range of architectures of different complexities that are able to serve as candidate end-to-end functions.

\paragraph{Single Layer Perceptron (SLP)} We consider the simplest case of a fully connected neural network with a single hidden layer that computes a linear combination of the input features prior to activation:
\begin{equation}
\label{eqn:slp}
\mathbf{X}_t = f(\mathbf{u}_t ; \boldsymbol{\theta}) = g(\mathbf{W}^\top \mathbf{u}_t + \mathbf{b})
\end{equation}
where $\mathbf{W} \in \mathbb{R}^{m \times N^t}$, $\mathbf{u}_t \in \mathbb{R}^{m} $ with $m = N^t \cdot \tau \cdot d$, $\mathbf{b} \in \mathbb{R}^{N^t}$and $g = \tanh$ is the activation function.

\paragraph{Multilayer Perceptron (MLP)} We consider a fully connected neural network with two hidden layers, representing a step up in model complexity from the SLP model:
\begin{equation}
\label{eqn:mlp}
\mathbf{X}_t = f(\mathbf{u}_t ; \boldsymbol{\theta}) = g \lbrack \mathbf{W}^{\lbrack 2 \rbrack\top} \sigma (\mathbf{W}^{\lbrack 1 \rbrack\top}\mathbf{u}_t + \mathbf{b}^{\lbrack 1 \rbrack}) + \mathbf{b}^{\lbrack 2 \rbrack} \rbrack
\end{equation}
where $\mathbf{W}^{\lbrack l \rbrack}$ and $\mathbf{b}^{\lbrack l \rbrack}$ are the weights and biases of the hidden layers and $g = \sigma = \tanh$ are the activation functions. Given the non-linear activation function $\sigma$, the model incorporates non-linearity with respect to the features of $\mathbf{u}_t$ prior to mapping into trading signals.

\paragraph{Convolutional Neural Networks (CNN)} Apart from their applications in computer vision \cite{simonyan2014very, he2016deep}, CNNs have been used to model multivariate time-series by incorporating convolutional filters that extract features from temporal data \cite{binkowski2018autoregressive}. These models rely on autoregressive causal convolutions and optionally, dilated convolutions \cite{oord2016wavenet} that allow the encoding of features over various receptive fields. The use of causal convolutions preserves the autoregressive ordering of temporal features, in that predictions made at time $t$ cannot depend on any future time steps $t+1, \cdots, T$.

\paragraph{Long Short-term Memory (LSTM)} Given the issues with exploding and (primarily) vanishing gradients in learning neural networks especially with long sequential data \cite{bengio1994learning}, RNNs that incorporate memory cell states and gating mechanisms such as LSTMs \cite{hochreiter1997long} have been used to address these limitations \cite{lim2021time}. The combination of memory cell states and gating functions helps in backpropagating gradients through time, allowing RNNs to better learn long-range dependencies in sequences.

We consider both the CNN and LSTM architectures for modelling spatio-temporal momentum. For full details of the implementations, we refer the reader to Appendix \ref{appendix_a}.

\subsection{Training Details}
\subsubsection{Loss Function}
Incorporating risk-adjusted metrics such as the Sharpe ratio during optimization have allowed models to better incorporate risk into learning and generation of trading signals \cite{lim2019enhancing}. Given a set $\mathcal{D} = \{(\mathbf{u}_t,~ \mathbf{X}_t = f(\mathbf{u}_t ; \boldsymbol{\theta})) \mid \mathbf{X}_t \in  [-1, 1]^{N^t}\}^T_{t=1}$ of spatio-temporal tensors and their corresponding signals, we define the loss function $\mathcal{L}_{\text{sharpe}}(\boldsymbol{\theta})$ over $\mathcal{D}$ as the annualized Sharpe ratio:
\begin{align}
\mathcal{L}_{\text{sharpe}}(\boldsymbol{\theta}) &= \sum\limits_{i=1}^{N^t} \lambda_i \cdot \mathcal{L}_{\text{sharpe}}^{(i)}(\boldsymbol{\theta}) \label{eqn:overall_sharpe_loss} \\
\mathcal{L}_{\text{sharpe}}^{(i)}(\boldsymbol{\theta}) &= - \frac{\sum_{t=1}^T R_i(t) \times \sqrt{252}}{\sum_{t=1}^T R_i(t)^2 - \left[\sum_{t=1}^T R_i(t)\right]^2} \\[2.5mm]
R_i(t) &= X_t^{(i)}~\frac{\sigma_{\text{tgt}}}{\sigma_t^{(i)}}~r_{t,t+1}^{(i)}
\end{align}
where $R_i(t)$ represents the volatility-scaled captured returns for asset $i$ from time $t$ to $t+1$. The multitask loss weights $\lambda_i$ balance task importance for the individual task-specific loss functions $\mathcal{L}_{\text{sharpe}}^{(i)}(\boldsymbol{\theta})$. In the case where all tasks (assets) are treated equally, we have $\lambda_i = 1 \slash N^t$.
\paragraph{Shrinkage Penalty} Taking into account the high dimensionality of the weight matrix $\mathbf{W}$ as per Equation (\ref{eqn:slp}), we incorporate $L_1$ regularization as an additional penalty term to the loss function of the SLP. This encourages feature selection and model sparsity as the weight coefficients of features that are of less relevance to prediction are shrunk towards zero. We incorporate $L_1$ regularization in the form of a penalty term on the sum of absolute weights of matrix $\mathbf{W}$ in the overall loss function:
\begin{align}
\mathcal{L}(\boldsymbol{\theta}) &= \mathcal{L}_{\text{sharpe}}(\boldsymbol{\theta}) + \alpha \sum_{w_{ij} \in \mathbf{W}} \lvert w_{ij} \rvert
\end{align}
where $\alpha$ is a hyperparameter controlling the shrinkage penalty term.

\subsubsection{Optimization}
In processing our datasets, we conduct a train-validation split with the earlier $90\%$ of data for training and the most recent $ 10\%$ reserved for validation. To minimize the overall empirical loss, we perform backpropagation using minibatch stochastic gradient descent with the Adam optimizer \cite{kingma2014adam}, and initiate early stopping based on the validation set loss. For all machine learning methods, we conduct hyperparameter optimization with 100 iterations of random search for selecting optimal candidate models. We refer the reader to Appendix \ref{appendix_a} for a detailed description of training parameters and specific details for the models.

\section{Performance Evaluation}
\label{performance_evaluation}

\subsection{Overview of Datasets}

\paragraph{US Equities} Our first dataset consists of 46 actively-traded US equities from the Financials sector with data obtained from the Center for Research in Security Prices (CRSP) \cite{crspdailystock}. We work with daily returns data, rebalancing the portfolio daily and performing strategy backtesting from 1990 to 2022.
\paragraph{Equity Index Futures} Our second dataset comprises 12 ratio-adjusted continuous equity index futures contracts with data obtained from the Pinnacle Data Corp CLC Database \cite{pinnacledata}. We work with daily returns data, performing portfolio rebalancing daily and backtesting from 2003 to 2020.

The full list of instruments is detailed in Appendix \ref{appendix_b}.

\subsection{Backtest Details}
We utilize an expanding window approach, where all models are trained with every iteration of 5 additional years as it becomes available. Taking US Equities as an example, the first iteration would involve training and validating on the period from 1990 to 1995, then fixing the model weights and evaluating the model on out-of-sample data from 1995 to 2000. The second iteration would involve training and validating from 1990 to 2000, and evaluating from 2000 to 2005, and so on. We conduct each experiment over multiple random seeded runs and report the aggregate performance.

\subsection{Momentum Features}
In constructing a general input spatio-temporal tensor $\mathbf{u}_t \in \mathbb{R}^{N^t \times \tau \times d}$ as per Equation (\ref{eqn:input_to_signal}), we include the below $d$ momentum features:
\begin{enumerate}[label = \textbf{F.\arabic*}]
    \item \textbf{Volatility Normalized Returns} -- we use $r^{(i)}_{t-k, t} \slash (\sigma^{(i)}_t \sqrt{k})$, representing asset returns normalized by daily volatility estimates scaled to a time scale $k \in \{1, 20, 63, 126, 252\}$, corresponding to daily, monthly, quarterly, semiannual and annual returns. 
    \item \textbf{MACD} -- we take volatility normalised MACD signals $Y_t^{(i)} (S_k, L_k)$ as per Equation (\ref{eqn:macd}) as input features, using short and long time scales $S_{k} \in \{8, 16, 32\}$ and $L_{k} \in \{24, 48, 96\}$.
\end{enumerate}
\subsection{Results and Discussion}
\label{section:results}
In evaluating the performance of all strategies, we use the following annualized metrics:
\begin{enumerate}[label = \textbf{M.\arabic*}]
    \item \textbf{Profitability} -- Expected Returns ($\mathbb{E}[\text{Returns}]$), Hit Rate
    \item \textbf{Risk} -- Volatility (Vol.), Downside Deviation, Maximum Drawdown (MDD)
    \item \textbf{Performance Ratios} -- Sharpe, Sortino and Calmar Ratios, Average Profit over Loss $\left(\frac{\text{Ave. P}}{\text{Ave. L}} \right)$
\end{enumerate}

For US Equities (and respectively for Equity Index Futures), we report the aggregated out-of-sample performance of all strategies from 1995 to 2022 (2008 to 2020) for overall returns computed in accordance with Equation (\ref{eqn:tsmom}). In this section, we compute the performance of the strategies in the absence of transaction costs to understand the raw predictive ability of the strategies. We provide an analysis of the impact of transaction costs in Section \ref{section:transaction_cost_impact}. We first present the performance of all strategies from their raw signal outputs in Table \ref{table:performance_raw} (Table \ref{table:performance_raw_pinnacle}). In order to facilitate the comparison between different strategies, we apply to all strategies an additional layer of volatility scaling at the portfolio level to target an annualized volatility of $15\%$ and report the performance in Table \ref{table:performance_portfolioscaled} (Table \ref{table:performance_portfolioscaled_pinnacle}) and cumulative returns in Figure \ref{fig:cumulative_returns_portfolio_level_scaling} (Figure \ref{fig:cumulative_returns_portfolio_level_scaling_pinnacle}).

\subsubsection{US Equities}
From Table \ref{table:performance_raw}, we first observe that the cross-sectional decile portfolio was an unprofitable strategy as seen from the CSMOM strategy delivering negative returns over the backtest period. In addition, classical time-series momentum portfolios like TSMOM and MACD generally underperformed compared to a simple long only approach. Two machine learning methods, the DMN and SLP were able to outperform the long only approach as seen from their performance ratios.

\begin{figure}[htbp]
\centering
\includegraphics[width=1.0\linewidth]{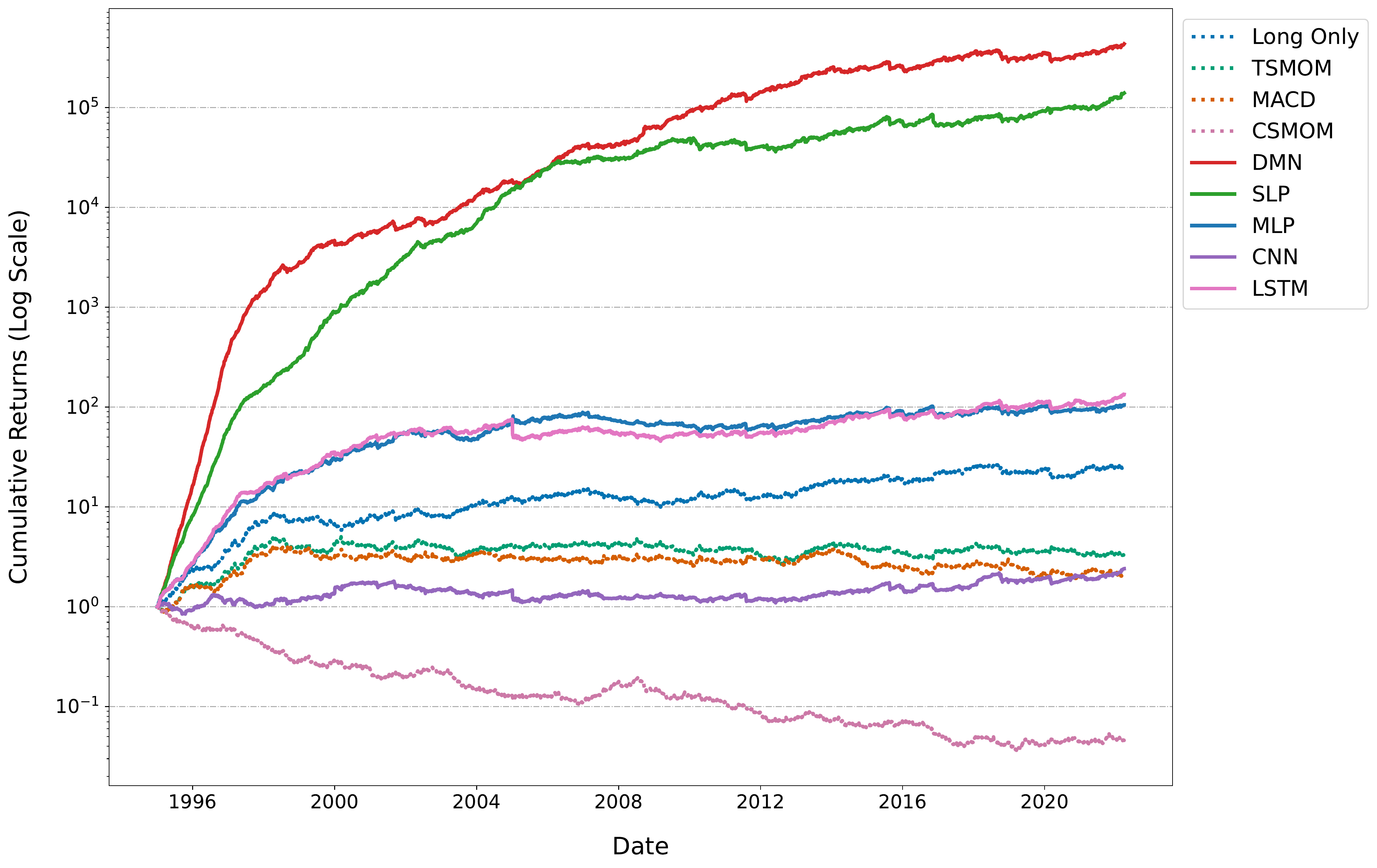}
\vspace{-0.3cm}
\caption{Cumulative Returns - Rescaled to Target Volatility \textbf{(US Equities)}}
\label{fig:cumulative_returns_portfolio_level_scaling}
\end{figure}

With the introduction of volatility scaling at the portfolio level, we observe improvements in performance across all strategies as shown in Table \ref{table:performance_portfolioscaled}. In particular, we see that spatio-temporal momentum (STMOM) methods, with the exception of CNN, experienced the largest increase in performance ratios among machine learning methods, with the Sharpe ratio of the SLP increasing by about 134\% (MLP - 251\%, LSTM - 217\%) as compared to an increase of 43\% for the DMN. This increase led to the MLP and LSTM outperforming the long only strategy. Both the DMN and SLP demonstrated the best performance and exhibited a large gap above all other methods as seen from Table \ref{table:performance_portfolioscaled} and the cumulative returns plot in Figure \ref{fig:cumulative_returns_portfolio_level_scaling}. This can be attributed to the DMN and SLP both capturing higher expected returns while being subject to lower downside risks. 

Within STMOM models, we observe a deterioration in performance of the STMOM strategy with an increased level of model complexity, as seen from the underperformance of the MLP, LSTM and CNN relative to a simple SLP model trained with a shrinkage penalty. This is contrary to the expectation that a model of greater complexity would be better suited to model the dynamics of spatio-temporal momentum. It is also possible that the underperformance in complex architectures is associated with difficulties in training models with multiple model configurations, especially the CNN as seen from its poor performance, echoing a similar conclusion from \cite{lim2019enhancing}. 

The underperformance of some STMOM models as compared to the DMN may be attributed to the relatively low signal-to-noise ratio of returns data, coupled with the limited amount of data available to STMOM strategies. During training, STMOM deep learners are exposed to only $t$ samples, while deep TSMOM strategies like the DMN have access to a much larger pool of $t \times N^t$ samples. As a result, overly complex STMOM models also run the risks of overfitting to in-sample data. 

\begin{table}[htbp]
\centering
\caption{Performance Metrics for Strategies -- Raw Signal Outputs \textbf{(US Equities)}}
\label{table:performance_raw}
\resizebox{\textwidth}{!}{
\begin{tabular}{lccccccccc}
\hline \toprule
& \textbf{E[Return]} & \textbf{Vol.} & \textbf{\begin{tabular}[c]{@{}l@{}} Downside \\ Deviation \end{tabular}} & \textbf{MDD} & \textbf{Sharpe} & \textbf{Sortino} & \textbf{Calmar} & \textbf{\begin{tabular}[c]{@{}l@{}} Hit \\ Rate \end{tabular}} & \textbf{$\mathbf{\frac{\text{Ave. P}}{\text{Ave. L}}}$} \\ 
\midrule
{\underline{\textbf{Benchmarks}}} &   &   &   &   &   &   &   &   &  \\
Long Only                 & \textbf{0.068} &                 0.102 &              0.072 &            0.235 &        0.667 &         0.944 &         0.290 &    0.541 &     0.951                                                    \\
TSMOM             & 0.012 &                 0.067 &               0.050 &            0.287 &        0.177 &          0.240 &        0.042 &    0.526 &     0.932                                                    \\
MACD                      & 0.001 &                 0.045 &              0.033 &            0.237 &        0.021 &         0.029 &        0.004 &    0.519 &     0.931                                                    \\ 
CSMOM                     & -0.033 &                 0.048 &              0.036 &            0.631 &       -0.702 &        -0.937 &       -0.053 &    0.494 &     0.911                                                    \\
\midrule
{\underline{\textbf{Reference}}} &   &   &   &   &   &   &   &   &  \\
DMN                    & 0.056 &                 \textbf{0.028} &              \textbf{0.018} &            \textbf{0.067} &        \textbf{2.043} &         \textbf{3.145} &        \textbf{0.924} &    \textbf{0.598} &     \textbf{1.055}                                                    \\
                    & (0.008) &               (0.008) &            (0.005) &          (0.029) &      (0.263) &       (0.452) &      (0.258) &  (0.005) &   (0.054)                                                    \\
\midrule
{\underline{\textbf{STMOM}}}        &                    &                 &                                                                           &                 &                 &                  &                 &                                                                             &                                                          \\
SLP                    & 0.032 &                 0.029 &               0.020 &            0.075 &        1.114 &         1.609 &        0.435 &    0.581 &     0.955                                                    \\
                    & (0.006) &               (0.005) &            (0.004) &          (0.015) &      (0.182) &       (0.288) &      (0.092) &  (0.007) &   (0.041)                                                    \\

MLP              & 0.013 &                 0.044 &              0.032 &            0.176 &        0.296 &          0.410 &        0.082 &    0.544 &     0.902                                                    \\
                    & (0.005) &               (0.008) &            (0.006) &          (0.044) &      (0.106) &       (0.148) &       (0.040) &  (0.007) &   (0.038)                                                   \\
CNN                       & 0.010 &                 0.051 &              0.037 &            0.167 &        0.195 &         0.273 &        0.066 &    0.516 &      0.980                                                    \\
                    & (0.006) &               (0.006) &            (0.005) &          (0.037) &      (0.106) &        (0.150) &      (0.043) &  (0.007) &   (0.029)                                                   \\
LSTM                    & 0.014 &                 0.044 &              0.031 &            0.159 &         0.320 &         0.458 &        0.103 &    0.546 &     0.903                                                    \\ 
                    & (0.004) &               (0.008) &            (0.006) &          (0.053) &      (0.108) &       (0.158) &      (0.061) &  (0.009) &   (0.036)                                                   \\
\bottomrule \hline

\end{tabular}
}
\begin{flushleft}$_{\text{\ \ \ \ (Standard deviation shown in parentheses)}}$\end{flushleft}
\vfill
\end{table}
\begin{table}[htbp]
\centering
\caption{Performance Metrics for Strategies -- Rescaled to Target Volatility \textbf{(US Equities)}}
\label{table:performance_portfolioscaled}
\resizebox{\textwidth}{!}{
\begin{tabular}{lccccccccc}
\hline \toprule
& \textbf{E[Return]} & \textbf{Vol.} & \textbf{\begin{tabular}[c]{@{}l@{}} Downside \\ Deviation \end{tabular}} & \textbf{MDD} & \textbf{Sharpe} & \textbf{Sortino} & \textbf{Calmar} & \textbf{\begin{tabular}[c]{@{}l@{}} Hit \\ Rate \end{tabular}} & \textbf{$\mathbf{\frac{\text{Ave. P}}{\text{Ave. L}}}$} \\ 
\midrule
{\underline{\textbf{Benchmarks}}} &   &   &   &   &   &   &   &   &  \\
Long Only                 & 0.131 &                 0.155 &              0.109 &            0.344 &        0.841 &         1.197 &         0.380 &    0.541 &     0.976                                                    \\
TSMOM             & 0.056 &                 0.157 &              0.112 &             0.470 &        0.358 &         0.501 &        0.119 &    0.526 &      0.960                                                    \\
MACD                      & 0.038 &                 0.157 &              0.112 &            0.524 &        0.245 &         0.343 &        0.073 &    0.519 &     0.968                                                    \\ 
CSMOM                     & -0.101 &                 \textbf{0.154} &              0.115 &            0.964 &       -0.655 &         -0.880 &       -0.105 &    0.494 &     0.919                                                    \\
\midrule
{\underline{\textbf{Reference}}} &   &   &   &   &   &   &   &   &  \\
DMN                    & \textbf{0.487} &                 0.167 &              0.105 &             \textbf{0.260} &         \textbf{2.920} &         \textbf{4.647} &        \textbf{1.887} &    \textbf{0.598} &     \textbf{1.179}                                                    \\
                    & (0.019) &               (0.001) &            (0.002) &          (0.029) &      (0.119) &       (0.229) &      (0.181) &  (0.005) &    (0.020)                                                    \\
\midrule
{\underline{\textbf{STMOM}}}        &                    &                 &                                                                           &                 &                 &                  &                 &                                                                             &                                                          \\
SLP                    & 0.423 &                 0.162 &              \textbf{0.102} &            0.301 &        2.609 &         4.161 &        1.428 &    0.581 &      1.170                                                    \\
                    & (0.048) &               (0.001) &            (0.002) &          (0.033) &      (0.282) &       (0.524) &      (0.252) &  (0.007) &   (0.034)                                                    \\

MLP              & 0.175 &                 0.169 &               0.110 &             0.410 &         1.040 &          1.590 &        0.439 &    0.544 &     1.032                                                    \\
                    & (0.030) &               (0.013) &            (0.002) &           (0.070) &      (0.178) &        (0.270) &      (0.109) &  (0.007) &    (0.020)                                                   \\
CNN                       & 0.035 &                   0.200 &              0.139 &            0.665 &        0.192 &         0.314 &        0.079 &    0.516 &     0.977                                                    \\
                    & (0.041) &               (0.059) &            (0.056) &          (0.318) &      (0.223) &        (0.340) &      (0.082) &  (0.007) &   (0.032)                                                    \\
LSTM                    & 0.178 &                 0.182 &              0.136 &            0.544 &        1.015 &         1.422 &        0.405 &    0.546 &     1.025                                                    \\ 
                    & (0.040) &               (0.028) &            (0.035) &          (0.264) &      (0.305) &       (0.511) &      (0.202) &  (0.009) &   (0.028)                                                   \\
\bottomrule \hline

\end{tabular}
}
\vfill
\end{table}

\subsubsection{Equity Index Futures}
Similar to US Equities, it is clear from Table \ref{table:performance_raw_pinnacle} that the cross-sectional decile portfolio was again unprofitable with the CSMOM strategy delivering negative returns. Both classical time-series momentum portfolios like TSMOM and MACD and the reference DMN were profitable but underperformed a long only approach. With the exception of CNN, all STMOM methods demonstrated the best performance, with the SLP and MLP models outperforming the DMN by more than four times as seen from their performance ratios.

\begin{figure}[htbp]
\centering
\includegraphics[width=1.0\linewidth]{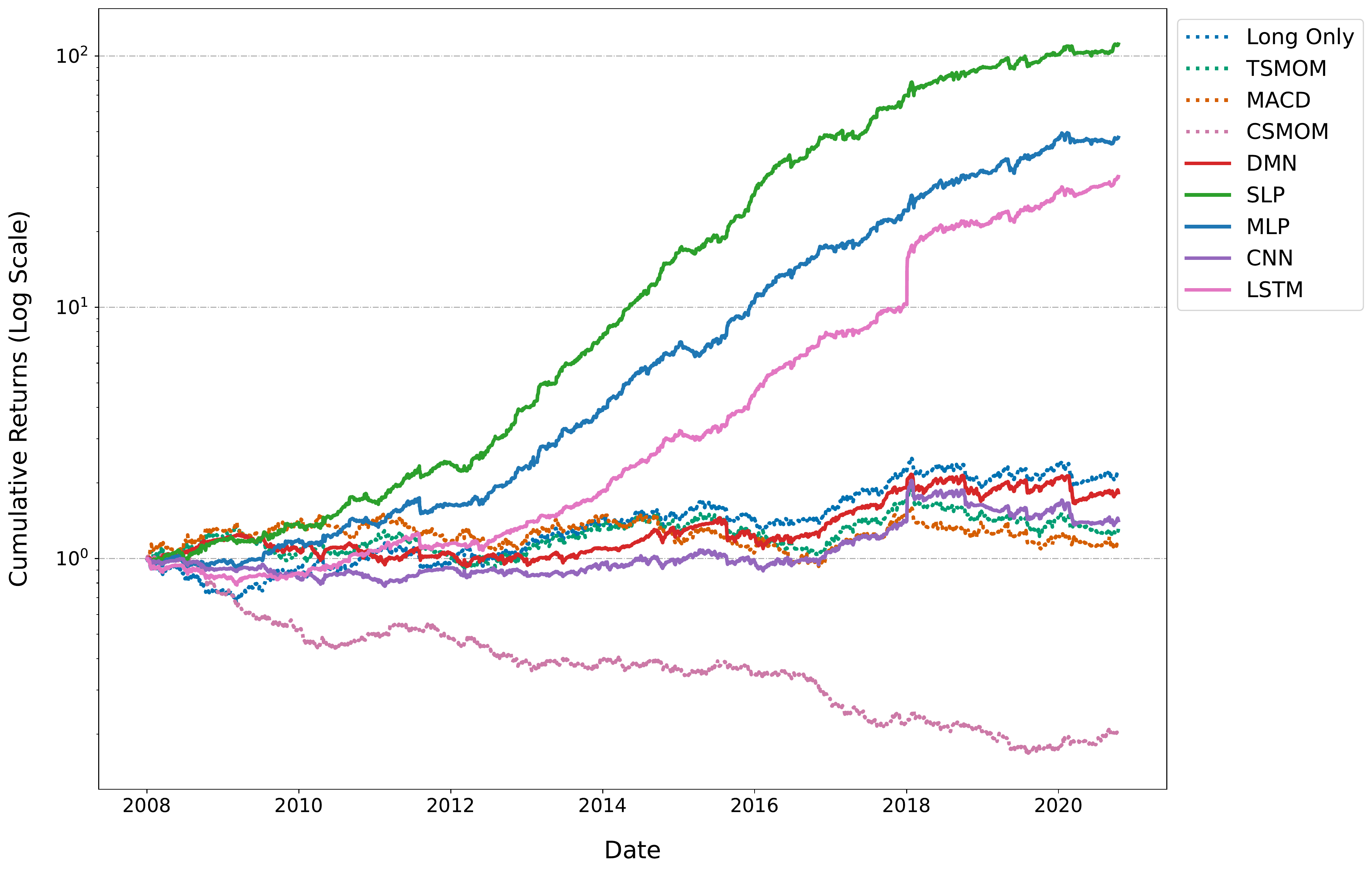}
\vspace{-0.3cm}
\caption{Cumulative Returns - Rescaled to Target Volatility \textbf{(Equity Index Futures)}}
\label{fig:cumulative_returns_portfolio_level_scaling_pinnacle}
\end{figure}

Incorporating volatility scaling at the portfolio level, we notice improvements in the performance for all strategies as shown in Table \ref{table:performance_portfolioscaled_pinnacle}. The increase in performance ratios for the benchmarks strategies and DMN were minimal, whereas STMOM methods experienced larger increases in their performance ratios with volatility scaling. This increase drove STMOM strategies to further outperform all other strategies, with the SLP model outperforming the DMN by more than six times in risk-adjusted returns. Echoing the observation for US Equities, the simplest SLP model again demonstrated the best performance above all other STMOM methods as seen from Table \ref{table:performance_portfolioscaled_pinnacle} and the cumulative returns plot in Figure \ref{fig:cumulative_returns_portfolio_level_scaling_pinnacle}, further lending support to the notion that a model of lower complexity is better suited to model spatio-temporal momentum.

\begin{table}[htbp]
\centering
\caption{Performance Metrics for Strategies -- Raw Signal Outputs \textbf{(Equity Index Futures)}}
\label{table:performance_raw_pinnacle}
\resizebox{\textwidth}{!}{
\begin{tabular}{lccccccccc}
\hline \toprule
& \textbf{E[Return]} & \textbf{Vol.} & \textbf{\begin{tabular}[c]{@{}l@{}} Downside \\ Deviation \end{tabular}} & \textbf{MDD} & \textbf{Sharpe} & \textbf{Sortino} & \textbf{Calmar} & \textbf{\begin{tabular}[c]{@{}l@{}} Hit \\ Rate \end{tabular}} & \textbf{$\mathbf{\frac{\text{Ave. P}}{\text{Ave. L}}}$} \\ 
\midrule
{\underline{\textbf{Benchmarks}}} &   &   &   &   &   &   &   &   &  \\
Long Only                 & \textbf{0.054} &                 0.125 &              0.093 &            0.275 &        0.427 &         0.574 &        0.195 &    0.549 &     0.883                                                    \\
TSMOM             & 0.020 &                 0.107 &              0.078 &            0.251 &        0.191 &         0.262 &        0.081 &    0.523 &     0.944                                                    \\
MACD                      & 0.006 &                 0.068 &               0.050 &            0.175 &        0.081 &         0.111 &        0.032 &    0.529 &     0.905                                                    \\ 
CSMOM                     & -0.048 &                 0.081 &               0.060 &            0.554 &       -0.584 &        -0.786 &       -0.086 &    0.489 &     0.946                                                    \\
\midrule
{\underline{\textbf{Reference}}} &   &   &   &   &   &   &   &   &  \\
DMN                    & 0.012 &                 0.034 &              0.026 &            0.102 &        0.301 &          0.410 &         0.110 &    0.525 &     0.971                                                    \\
                    & (0.017) &               (0.024) &            (0.018) &          (0.066) &       (0.220) &       (0.307) &      (0.088) &  (0.009) &   (0.042)                                                    \\
\midrule
{\underline{\textbf{STMOM}}}        &                    &                 &                                                                           &                 &                 &                  &                 &                                                                             &                                                          \\
SLP                    & 0.047 &                 0.049 &              0.035 &            0.152 &        1.242 &         1.856 &        0.657 &    \textbf{0.574} &     0.992                                                    \\
                    & (0.017) &               (0.019) &            (0.015) &          (0.081) &      (0.834) &       (1.433) &      (0.781) &  (0.018) &   (0.142)                                                    \\

MLP              & 0.037 &                 \textbf{0.032} &              \textbf{0.022} &            \textbf{0.088} &        \textbf{1.319} &          \textbf{2.050} &        \textbf{0.683} &    0.559 &     \textbf{1.075}                                                    \\
                    & (0.014) &               (0.013) &             (0.010) &          (0.063) &      (0.593) &       (1.065) &      (0.478) &  (0.014) &   (0.127)                                                   \\
CNN                       & -0.016 &                 0.065 &              0.049 &            0.291 &       -0.215 &        -0.283 &       -0.045 &    0.518 &     0.886                                                    \\
                    & (0.014) &                (0.020) &            (0.015) &          (0.111) &       (0.170) &       (0.228) &      (0.033) &  (0.014) &   (0.048)                                                   \\
LSTM                    & 0.039 &                 0.057 &               0.040 &            0.172 &        0.746 &         1.079 &        0.289 &     0.550 &     0.973                                                    \\ 
                    & (0.013) &               (0.014) &             (0.010) &           (0.070) &      (0.329) &       (0.529) &      (0.198) &  (0.016) &   (0.098)                                                   \\
\bottomrule \hline

\end{tabular}
}
\begin{flushleft}$_{\text{\ \ \ \ (Standard deviation shown in parentheses)}}$\end{flushleft}
\vfill
\end{table}
\begin{table}[htbp]
\centering
\caption{Performance Metrics for Strategies -- Rescaled to Target Volatility \textbf{(Equity Index Futures)}}
\label{table:performance_portfolioscaled_pinnacle}
\resizebox{\textwidth}{!}{
\begin{tabular}{lccccccccc}
\hline \toprule
& \textbf{E[Return]} & \textbf{Vol.} & \textbf{\begin{tabular}[c]{@{}l@{}} Downside \\ Deviation \end{tabular}} & \textbf{MDD} & \textbf{Sharpe} & \textbf{Sortino} & \textbf{Calmar} & \textbf{\begin{tabular}[c]{@{}l@{}} Hit \\ Rate \end{tabular}} & \textbf{$\mathbf{\frac{\text{Ave. P}}{\text{Ave. L}}}$} \\ 
\midrule
{\underline{\textbf{Benchmarks}}} &   &   &   &   &   &   &   &   &  \\
Long Only                 & 0.070 &                 \textbf{0.154} &              0.114 &            0.319 &        0.456 &         0.616 &        0.221 &    0.549 &     0.887                                                    \\
TSMOM             & 0.033 &                 \textbf{0.154} &              0.112 &            0.355 &        0.212 &         0.291 &        0.092 &    0.523 &     0.946                                                    \\
MACD                      & 0.023 &                 0.156 &              0.114 &             0.390 &        0.145 &         0.198 &        0.058 &    0.529 &     0.915                                                    \\ 
CSMOM                     & -0.107 &                 \textbf{0.154} &              0.115 &            0.838 &       -0.697 &        -0.929 &       -0.128 &    0.489 &     0.929                                                   \\
\midrule
{\underline{\textbf{Reference}}} &   &   &   &   &   &   &   &   &  \\
DMN                    & 0.055 &                 0.162 &              0.115 &            0.391 &         0.340 &         0.478 &        0.156 &    0.525 &     0.966                                                    \\
                    & (0.026) &               (0.006) &            (0.002) &          (0.086) &      (0.165) &       (0.233) &      (0.093) &  (0.009) &   (0.028)                                                    \\
\midrule
{\underline{\textbf{STMOM}}}        &                    &                 &                                                                           &                 &                 &                  &                 &                                                                             &                                                          \\
SLP                    & \textbf{0.333} &                 0.161 &              0.104 &            0.244 &        \textbf{2.066} &         \textbf{3.228} &        \textbf{1.619} &    \textbf{0.574} &     1.121                                                    \\
                    & (0.084) &               (0.003) &            (0.004) &          (0.076) &      (0.498) &        (0.910) &      (0.957) &  (0.018) &   (0.071)                                                    \\

MLP              & 0.288 &                 0.162 &              0.107 &            \textbf{0.238} &        1.776 &         2.699 &        1.302 &    0.559 &     \textbf{1.123}                                                    \\
                    & (0.058) &               (0.004) &            (0.005) &          (0.059) &      (0.333) &        (0.560) &      (0.452) &  (0.014) &   (0.048)                                                   \\
CNN                       & 0.030 &                 0.164 &              0.112 &            0.443 &        0.174 &          0.270 &        0.089 &    0.518 &     0.963                                                    \\
                    & (0.046) &               (0.015) &            (0.002) &          (0.121) &      (0.279) &       (0.412) &      (0.101) &  (0.014) &   (0.023)                                                    \\
LSTM                    & 0.251 &                 0.195 &              \textbf{0.103} &            0.298 &        1.389 &         2.451 &         0.890 &     0.550 &     1.118                                                    \\ 
                    & (0.085) &               (0.115) &            (0.004) &          (0.063) &      (0.384) &       (0.892) &      (0.415) &  (0.016) &   (0.087)                                                   \\
\bottomrule \hline

\end{tabular}
}
\vfill
\end{table}

For subsequent sections, we focus our analysis on the US Equities dataset.

\subsection{Signal Diversification}
\begin{figure}[htbp]
\begin{subfigure}{0.495\textwidth}
\includegraphics[width=\linewidth]{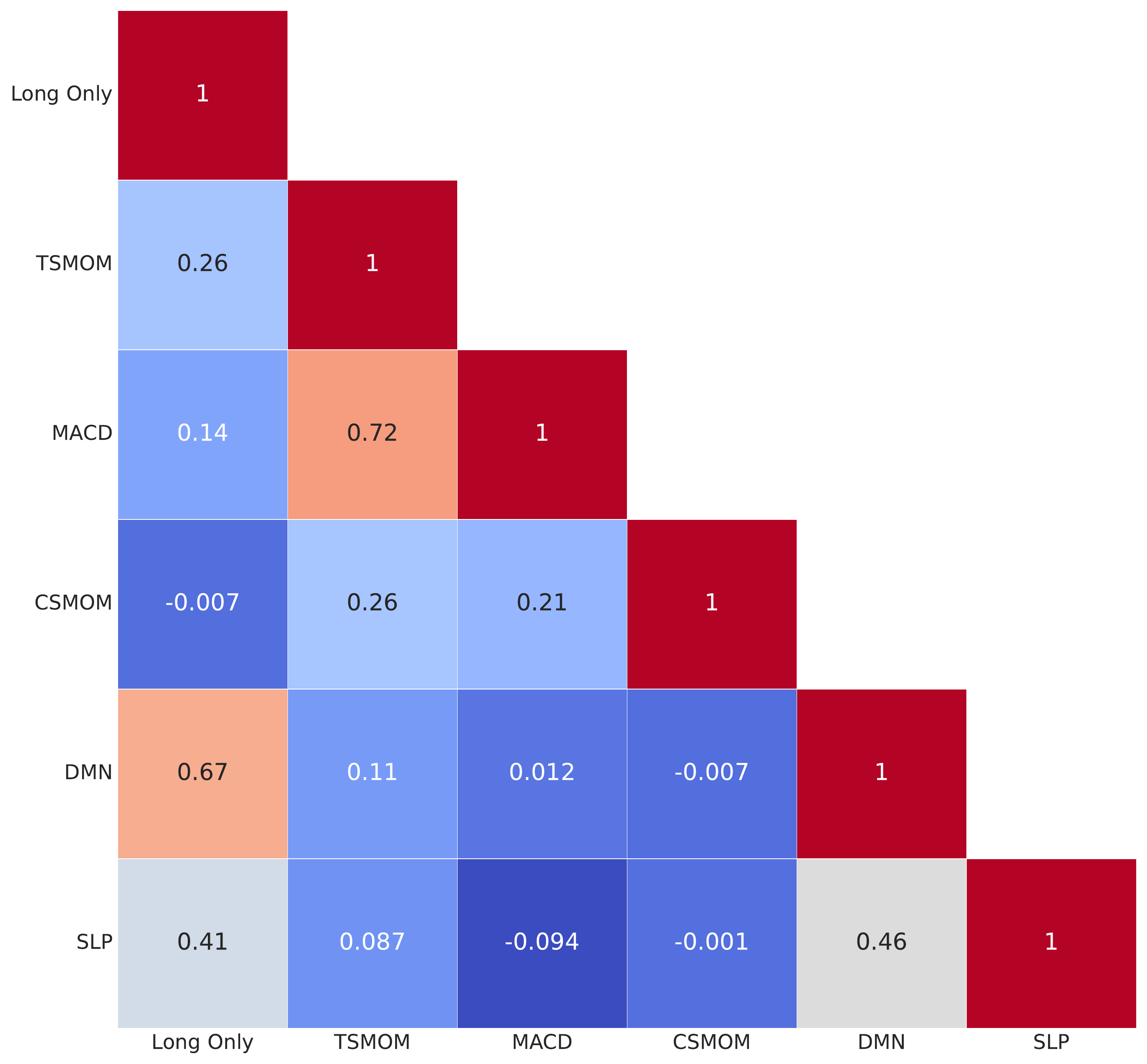} 

\end{subfigure}
\begin{subfigure}{0.495\textwidth}
\includegraphics[width=\linewidth, height=6.4cm]{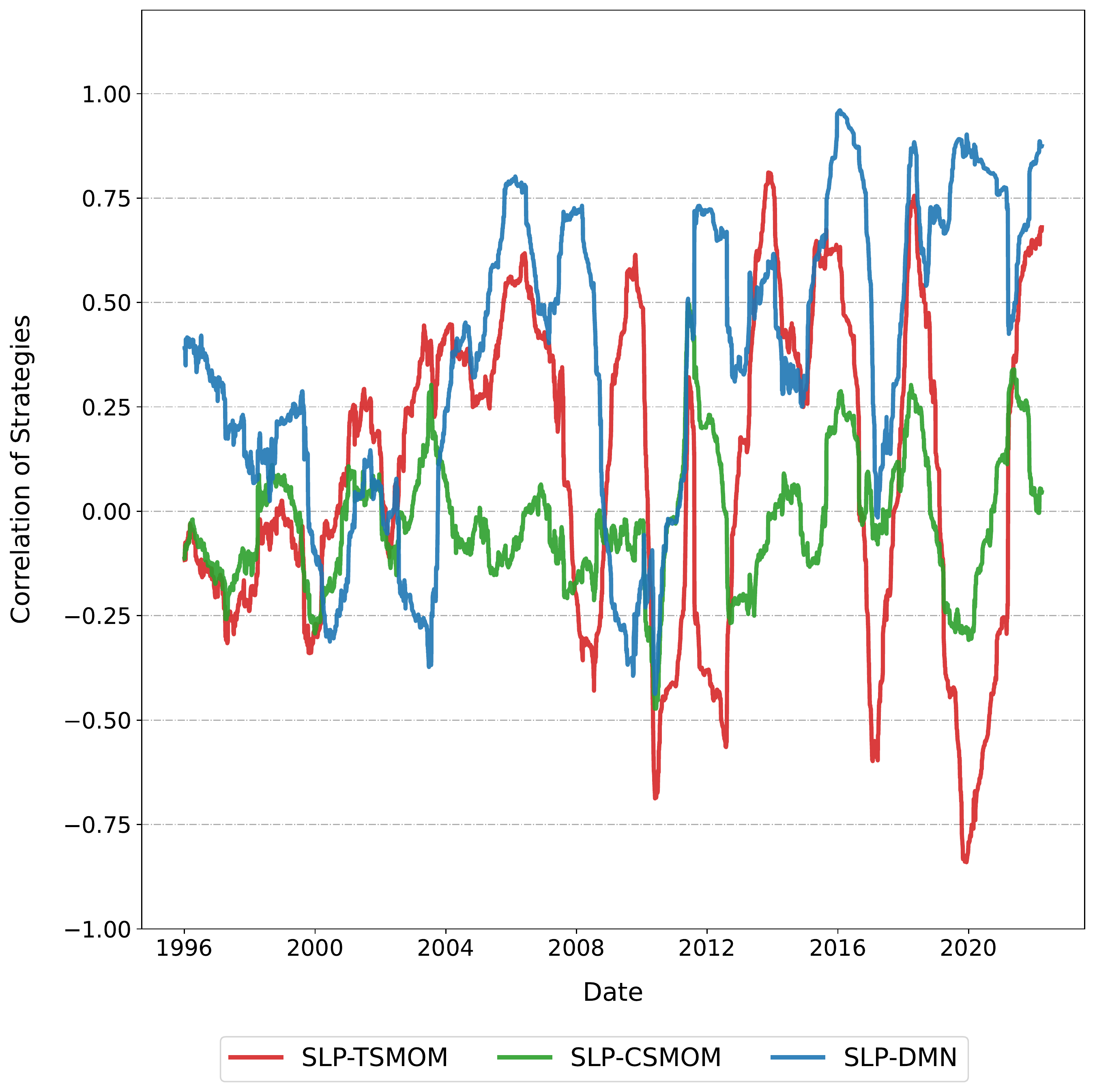}
\end{subfigure}
\caption{\textbf{(Left)} Correlation of Strategies \textbf{(Right)} 1Y Rolling Correlations of SLP \textbf{(US Equities)}}
\label{fig:correlation_matrix_rolling_correlation}
\end{figure}

From Figure \ref{fig:correlation_matrix_rolling_correlation}, the SLP exhibits mostly zero to moderately positive correlation with other momentum strategies, with the correlation between SLP and DMN at about $46\%$. The correlations of the SLP with other strategies are also fairly unstable, with the SLP periodically displaying a negative correlation with TSMOM, CSMOM and DMN, translating into possible benefits from signal diversification.

We evaluate the combination of time-series momentum (TSMOM and DMN) with a cross-sectional momentum strategy (CSMOM) and compare it to a spatio-temporal momentum strategy. Focusing on the Sharpe ratio, we see from Table \ref{table:performance_portfolioscaled_combination} that the combinations of both TSMOM+CSMOM (0.177) and DMN+CSMOM (2.115) portfolios do not outperform a single SLP portfolio (2.609), demonstrating that the benefit of using a spatio-temporal momentum strategy outweighs a portfolio that combines both time-series and cross-sectional momentum, thus indicating that the model is learning meaningful interactions between the time-series and cross-sectional domain which are not captured by separate models. In addition, we analyze the combination of a time-series momentum (DMN) with a spatio-temporal momentum strategy (SLP). The combination of DMN (2.920) and SLP (2.609) yielded an overall higher Sharpe ratio (3.304), highlighting substantial scope for strategy diversification by incorporating a spatio-temporal momentum strategy.

\begin{table}[htbp]
\centering
\caption{Performance Metrics for Strategy Combinations -- Rescaled to Target Volatility \textbf{(US Equities)}}
\label{table:performance_portfolioscaled_combination}
\resizebox{\textwidth}{!}{
\begin{tabular}{lccccccccc}
\hline \toprule
& \textbf{E[Return]} & \textbf{Vol.} & \textbf{\begin{tabular}[c]{@{}l@{}} Downside \\ Deviation \end{tabular}} & \textbf{MDD} & \textbf{Sharpe} & \textbf{Sortino} & \textbf{Calmar} & \textbf{\begin{tabular}[c]{@{}l@{}} Hit \\ Rate \end{tabular}} & \textbf{$\mathbf{\frac{\text{Ave. P}}{\text{Ave. L}}}$} \\ 
\midrule

{\underline{\textbf{Combination}}}        &                    &                 &                                                                           &                 &                 &                  &                 &                                                                             &                                                          \\
TSMOM+CSMOM                    & 0.028 &                  \textbf{0.156} &               0.113 &             0.577 &         0.177 &          0.245 &         0.048 &     0.521 &      0.946                                                    \\
DMN+CSMOM                    & 0.340 &                  0.161 &               0.106 &             0.307 &         2.115 &          3.222 &         1.132 &     0.572 &      1.099                                                    \\
                    & (0.025) &                  (0.001) &               (0.001) &             (0.045) &         (0.153) &          (0.261) &         (0.189) &     (0.005) &      (0.017)                                                    \\
DMN+SLP                    & \textbf{0.551} &                  0.167 &               \textbf{0.101} &             \textbf{0.230} &         \textbf{3.304} &          \textbf{5.441} &         \textbf{2.402} &     \textbf{0.603} &      \textbf{1.229}                                                    \\
                    & (0.027) &                  (0.001) &               (0.001) &             (0.013) &         (0.151) &          (0.308) &         (0.147) &     (0.005) &      (0.026)                                                    \\

\bottomrule \hline
\end{tabular}
}
\vfill
\end{table}

\subsection{Transaction Costs and Turnover Regularization}
\label{section:transaction_cost_impact}
\begin{figure}[htbp]
\centering
\includegraphics[width=1.0\linewidth]{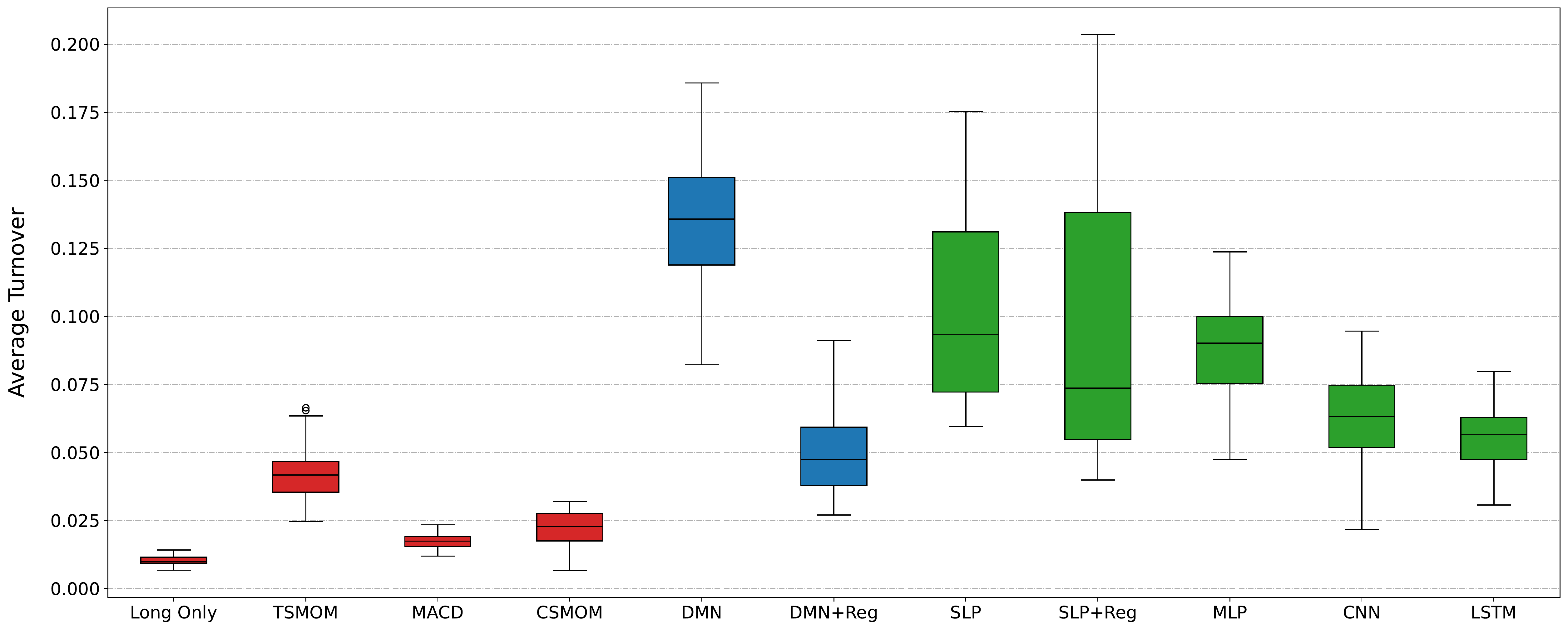}
\vspace{-0.3cm}
\caption{Average Turnover of Strategies \textbf{(US Equities)}}
\label{fig:average_turnover}
\end{figure}

In this section, we analyze the impact of transaction costs on the performance of all strategies. Following \cite{baltas2020demystifying}, we define turnover $\text{TO}_t^{(i)}$ as the absolute daily change in the trading signal of an asset:
\begin{equation}
\text{TO}_t^{(i)} = \sigma_{\text{tgt}} \left| \frac{X_t^{(i)}}{\sigma_t^{(i)}} -  \frac{X_{t-1}^{(i)}}{\sigma_{t-1}^{(i)}} \right|
\end{equation}
which encompasses the amount of rebalancing as determined by shifts in both volatility estimates and underlying positions. In Figure \ref{fig:average_turnover}, we plot the distributions of turnover averaged across all assets for individual strategies. We observe that machine learning models trade considerably more than the benchmarks, with the SLP having a lower turnover than the DMN. To study the impact of transactions costs on performance, we compute ex-cost Sharpe ratios using captured returns net of transaction costs $\tilde{r}_{t,t+1}^{\text{TSMOM}}$:
\begin{equation}
\label{eqn:returns_net_of_transaction_costs}
\tilde{r}_{t,t+1}^{\text{TSMOM}} = \frac{1}{N_t} \sum_{i=1}^{N_t} \bigg(X_t^{(i)}~ \frac{\sigma_{\text{tgt}}}{\sigma_t^{(i)}}~r_{t,t+1}^{(i)} -  c \cdot \text{TO}_t^{(i)} \bigg)
\end{equation}
where $c$ is a constant approximating the specific transaction cost scenario. Based on the ex-cost Sharpe Ratios in Table \ref{table:sharpe_ratio_by_transaction_cost}, the non-regularized DMN and SLP models are able to retain their performance over benchmarks for transactions costs of up to $c=5$ to $10$ basis points, which are considered to be realistic transaction costs for equity markets. Following the approach of \cite{lim2019enhancing}, we investigate the effect of incorporating turnover regularization during training in the form of optimizing for ex-cost Sharpe ratios computed using $\tilde{r}_{t,t+1}^{\text{TSMOM}}$. For non-recurrent models that do not generate sequential signals, such as the SLP, we introduce a localized minibatch turnover regularization in the form: 
\begin{equation}
\label{eqn:minibatch_turnover}
\widetilde{\text{TO}}_t^{(i)} = \sigma_{\text{tgt}} \left| \frac{X_t^{(i)}}{\sigma_t^{(i)}} -  \frac{X_{t^\ast}^{(i)}}{\sigma_{t^\ast}^{(i)}} \right|
\end{equation}
where $t \neq t^\ast$ with $t$ and $t^\ast$ representing consecutive samples within a given training minibatch. Compared to its non-regularized counterpart, incorporating turnover regularization for the DMN resulted in a lower turnover but better performance is only observed at a transaction cost of $c = 10$ basis points, while performance for all other transaction cost scenarios deteriorated. On the other hand, turnover regularization consistently improved the SLP's performance across all transaction cost scenarios. Referring to the distribution of average turnover for the regularized SLP, we observe a lower mean turnover but an increase in the spread between the minimum, maximum and interquartile range. This is likely a direct consequence of performing stochastic gradient descent with the localized minibatch turnover regularization as per Equation (\ref{eqn:minibatch_turnover}), requiring batching over a shuffled training set.

\begin{table}[htbp]
\centering
\caption{Impact of Transactions Costs on Sharpe Ratio -- Rescaled to Target Volatility \textbf{(US Equities)}}
\label{table:sharpe_ratio_by_transaction_cost}
\resizebox{\textwidth}{!}{
\begin{tabular}{lcccccccc}
\hline \toprule
\textbf{Transaction Cost (Basis Points)} & \textbf{0.0} & \textbf{0.5} & \textbf{1.0} & \textbf{2.0} & \textbf{3.0} & \textbf{4.0} & \textbf{5.0} & \textbf{10.0} \\ 
\midrule

{\underline{\textbf{Benchmarks}}}        &                    &                 &                                                                           &                 &                 &                  &                 &                                                                                                                                       \\
Long Only                    & 0.841 &  0.839 &  0.838 &  0.835 &  0.832 &  0.829 &  0.826 &  0.812                                                   \\
TSMOM                    & 0.358 &  0.347 &  0.336 &  0.315 &  0.293 &  0.271 &  0.249 &  0.140                                                   \\
MACD                    & 0.245 &  0.238 &  0.232 &  0.219 &  0.207 &  0.194 &  0.182 &  0.119                                                   \\
CSMOM                    & -0.655 & -0.683 & -0.710 & -0.765 & -0.820 & -0.875 & -0.930 & -1.204                                                   \\
\midrule
{\underline{\textbf{Reference}}}        &                    &                 &                                                                           &                 &                 &                  &                 &                                                                                                                                       \\
DMN                    & \textbf{2.920} &  \textbf{2.844} &  \textbf{2.768} &  \textbf{2.615} &  \textbf{2.462} &  \textbf{2.308} &  \textbf{2.153} &  1.375                                                   \\
DMN+Reg                    & 2.073 &  2.044 &  2.015 &  1.957 &  1.899 &  1.840 &  1.782 &  \textbf{1.486}                                                   \\

\midrule
{\underline{\textbf{STMOM}}}        &                    &                 &                                                                           &                 &                 &                  &                 &                                                                                                                                       \\
SLP                    & 2.609 &  2.518 &  2.427 &  2.243 &  2.060 &  1.876 &  1.691 &  0.762                                                   \\
SLP+Reg                    & \textbf{2.672} &  \textbf{2.603} &  \textbf{2.534} &  \textbf{2.395} &  \textbf{2.256} &  \textbf{2.116} &  \textbf{1.976} &  \textbf{1.271}                                                   \\
\bottomrule \hline
\end{tabular}
}
\vfill
\end{table}

\subsection{Interpretability} 

Apart from demonstrating the best performance, the SLP is the simplest model out of all STMOM architectures. We perform an analysis on the interpretability of the trained SLP using SHAP (SHapley Additive exPlanations) \cite{lundberg2017unified}, a model-agnostic interpretation method for computing feature importance and effects for machine learning models. SHAP values associated to each individual feature measures the change in the expected value of the model's prediction when conditioning on that feature. We approximate SHAP values with Deep SHAP \cite{lundberg2017unified}, a high-speed approximation algorithm which builds on DeepLIFT \cite{shrikumar2017learning}.

\begin{figure}[htbp]
\centering
\includegraphics[width=1.0\linewidth]{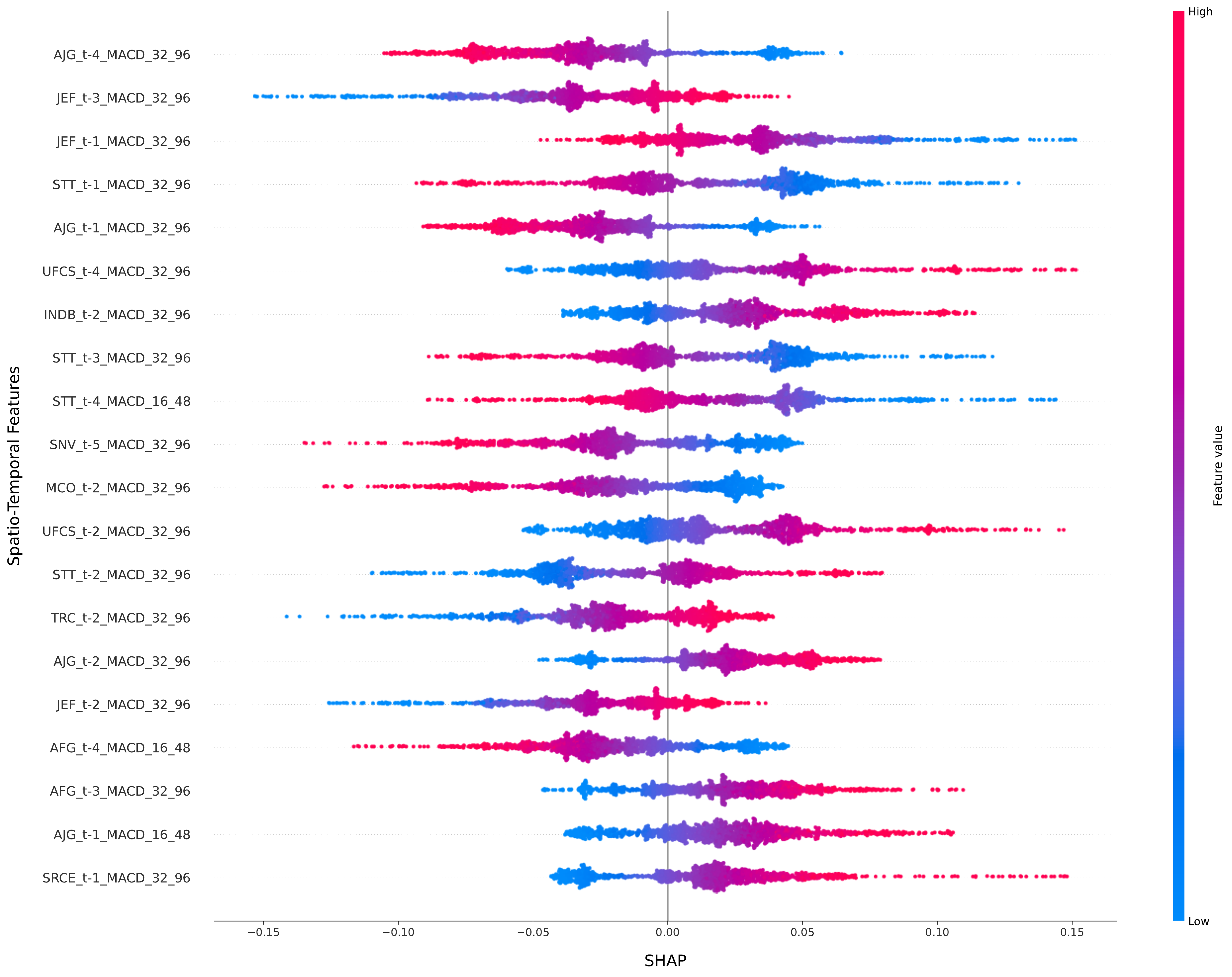}
\vspace{-0.3cm}
\caption{Impact on Predicted Signal Output for a Single Asset -- BAC \textbf{(US Equities)}}
\label{fig:shap_one_asset}
\end{figure}

\paragraph{Prediction for a Single Asset} We analyze the importance and effects of the spatio-temporal features on the predicted signal of a single asset -- BAC, by plotting SHAP values for the top 20 features ordered according to their importance in Figure \ref{fig:shap_one_asset}. It is evident that the top features are dominated by volatility normalized MACD, indicating that the MACD features of assets are, on average, contributing more to the predicted signal of BAC as compared to other types of features like volatility normalized returns. This could mean that the cross over of exponentially weighted moving averages are considered significant momentum features for the model in generating its trading signals. Interestingly, the top features do not contain much of the BAC’s own features, and instead other assets’ features in the cross-section are considered important in its own predicted signal.

Moving from low to high feature values and vice versa, there are clear patterns in the resultant impact of spatio-temporal features on the predicted signal output as seen from the gradual transitions in colour. Momentum and mean-reversion effects of individual features are also relatively distinct and can be inferred from the SHAP plot. Taking the first two features as examples, the first feature "AJG\_t-4\_MACD\_32\_96" exhibits a mean-reversion effect as higher feature values lead to a lower predicted signal for BAC, while the second feature "JEF\_t-3\_MACD\_32\_96" shows a momentum effect as higher feature values lead to a higher predicted signal for BAC.

\begin{figure}[htbp]
\centering
\includegraphics[width=1.0\linewidth]{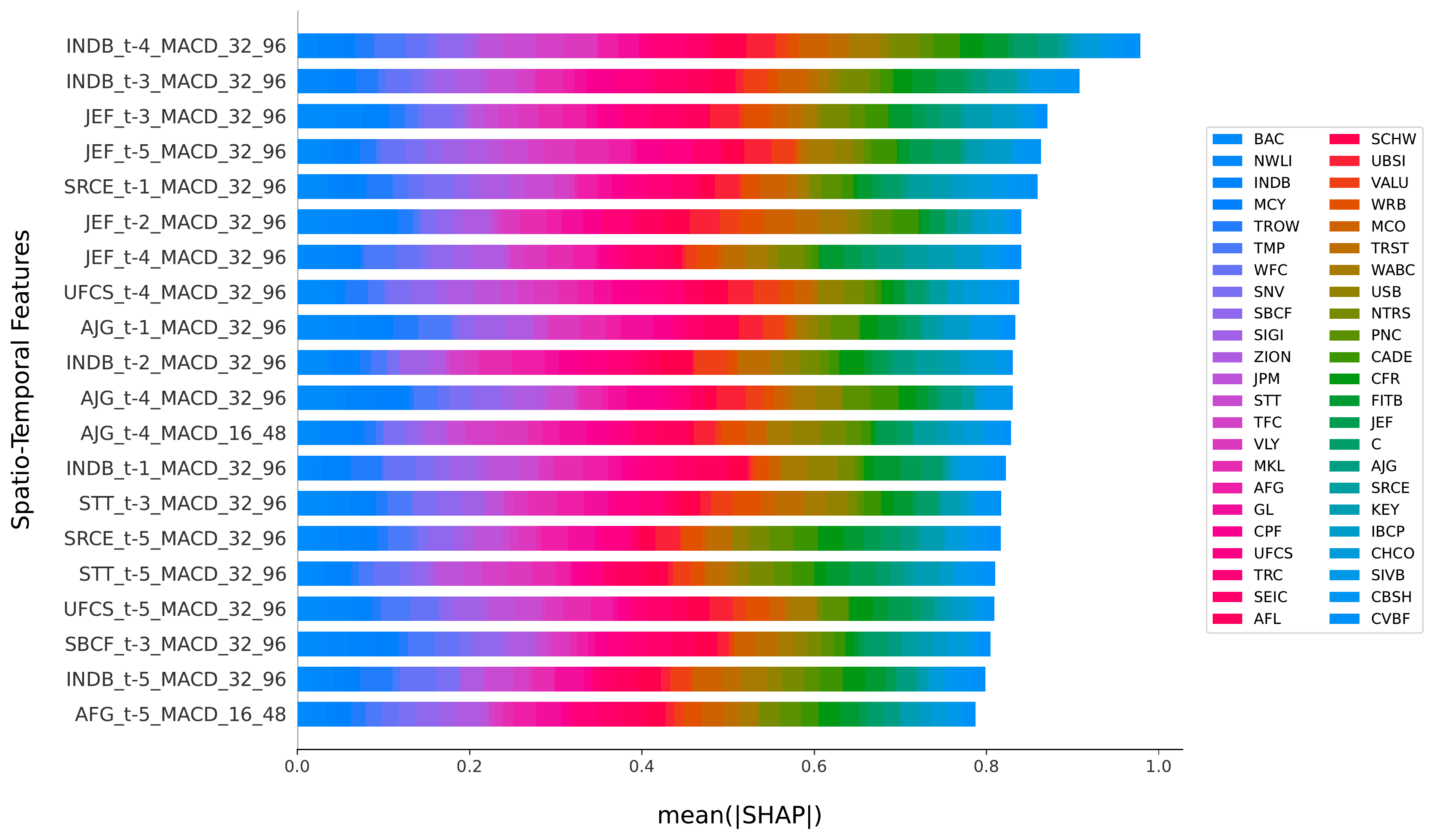}
\vspace{-0.3cm}
\caption{Cumulative Mean Absolute Impact on Predicted Signal Output for All Assets \textbf{(US Equities)}}
\label{fig:shap_all_assets}
\end{figure}

\paragraph{Prediction for All Assets} To examine global feature importance, we compute the mean absolute SHAP values for each spatio-temporal feature on the predicted signal for all assets in US Equities and plot their cumulative impact in Figure \ref{fig:shap_all_assets}. Ordered according to importance, the top 20 features are again dominated by volatility normalized MACD features. We note that the cumulative absolute impact of a feature does not distinguish between its momentum or mean-reversion effects, and serves only as a proxy of its feature importance over all assets.

\section{Conclusions}
\label{conclusion}
We introduce Spatio-Temporal Momentum, a class of machine learning models that combine time-series and cross-sectional momentum strategies. This class of momentum strategies simultaneously generates trading signals for a portfolio of assets in a multitask setting, using both time-series and cross-sectional features from all assets in the portfolio. We demonstrate that the SLP, a neural network with a single hidden layer trained with a shrinkage penalty, is able to outperform more complex models in modelling spatio-temporal momentum, demonstrating its effectiveness over benchmarks on two different datasets.

We show that the performance of a spatio-temporal momentum strategy can be superior to a simple blend of time-series and cross-sectional momentum strategies. Exhibiting unstable correlation with other momentum strategies, the spatio-temporal momentum strategy when combined with a time-series momentum strategy like the DMN yields higher risk-adjusted returns, making spatio-temporal momentum a valuable strategy for diversification.

Given various cost scenarios, we examine the impact of transaction costs on the performance of all strategies and show that both the DMN and SLP were able to retain their performance over benchmarks for costs up to 5 to 10 basis points when tested on the US equities data. We incorporate turnover regularization for the DMN and a different localized minibatch turnover regularization for non-recurrent models like the SLP, resulting in the SLP achieving better performance ratios across all transaction cost scenarios.

With the simplicity of the SLP model, it is possible to directly visualize and interpret its weights corresponding to each spatio-temporal momentum feature. We analyze the importance of individual spatio-temporal momentum features using the model-agnostic SHAP method, revealing clear patterns in momentum and mean-reversion effects of individual features for predicting the trading signal of an individual asset, as well as showing global feature importance for all assets.

In future works, we would like to study the performance of spatio-temporal momentum using alternative feature representations. Another extension of this work includes adapting attention-based deep learning architectures \cite{wood2021trading, bahdanau2014neural, vaswani2017attention} and investigating their effectiveness in modelling spatio-temporal momentum.

\bibliographystyle{plain}
\bibliography{bibliography}

\clearpage
\appendix

\section{Deep Learning Architectures \& Optimization}
\label{appendix_a}
\subsection{Deep Learning Architectures}
For the CNN and LSTM models, we consider the spatio-temporal tensor $\mathbf{u}_t \in \mathbb{R}^{\tau \times m^{\prime}}$ with $m^{\prime} = N^t \cdot d$, representing an input sequence of momentum features of all assets over a temporal history $\tau$.

\paragraph{Convolutional Neural Networks (CNN)} We consider a 1-D autoregressive CNN of the following:
\begin{align}
\mathbf{h}_t &= \mathcal{P} \sigma \lbrack \mathbf{W}_c^{\lbrack 2 \rbrack}\ast \sigma(\mathbf{W}_c^{\lbrack 1 \rbrack}\ast\mathbf{u}_t + \mathbf{b}_c^{\lbrack 1 \rbrack}) + \mathbf{b}_c^{\lbrack 2 \rbrack} \rbrack \\
\mathbf{X}_t &= f(\mathbf{u}_t ; \boldsymbol{\theta}) = g \lbrack \mathbf{W}^{\lbrack 2 \rbrack\top} \sigma (\mathbf{W}^{\lbrack 1 \rbrack\top}\mathbf{h}_t + \mathbf{b}^{\lbrack 1 \rbrack}) + \mathbf{b}^{\lbrack 2 \rbrack} \rbrack
\end{align}
where $\mathbf{W}_c^{\lbrack l \rbrack}$ represent convolutional kernels with associated bias terms $\mathbf{b}_c^{\lbrack l \rbrack}$ and activations $\sigma = \tanh$, and $\mathbf{w} \ast \mathbf{u}$ representing the causal convolution operation between the input sequence $\mathbf{u}$ and kernel $\mathbf{w}$. Additionally, we interface an average pooling layer $\mathcal{P}$ that performs a downsampling step prior to propagating the activations $\mathbf{h}_t$ into an MLP as per Equation (\ref{eqn:mlp}).

\paragraph{Long Short-term Memory (LSTM)} We consider a single layer LSTM model:
\begin{align}
\mathbf{\Gamma}^i_t &= \sigma_S(\mathbf{W}_{i} \mathbf{u}_t + \mathbf{V}_{i} \mathbf{h}_{t-1} + \mathbf{b}_i) \\
\mathbf{\Gamma}^f_t &= \sigma_S(\mathbf{W}_{f} \mathbf{u}_t + \mathbf{V}_{f} \mathbf{h}_{t-1} + \mathbf{b}_f) \\
\mathbf{\Gamma}^o_t &= \sigma_S(\mathbf{W}_{o} \mathbf{u}_t + \mathbf{V}_{o} \mathbf{h}_{t-1} + \mathbf{b}_o) \\
\mathbf{\tilde{c}}_t &= g(\mathbf{W}_{c} \mathbf{u}_t + \mathbf{V}_{c} \mathbf{h}_{t-1} + \mathbf{b}_c) \\
\mathbf{c}_t &= \mathbf{\Gamma}^i_t \odot \mathbf{\tilde{c}}_t + \mathbf{\Gamma}^f_t \odot \mathbf{c}_{t-1} \\
\mathbf{h}_t &= \mathbf{\Gamma}^o_t \odot g(\mathbf{c}_t) \\
\mathbf{X}_t &= g(\mathbf{W}_{\text{dist}} \mathbf{h}_t + \mathbf{b})
\end{align}
where $\mathbf{W}_\ast$, $\mathbf{V}_\ast$ represent the weights and $\mathbf{b}_\ast$ the biases of the input, forget and output gates $\mathbf{\Gamma}^i_t$, $\mathbf{\Gamma}^f_t$, $\mathbf{\Gamma}^o_t$ respectively, with activation functions $\sigma_S = \text{sigmoid}$ and $g = \tanh$. Subsequently, the LSTM computes the memory cell state $\mathbf{c}_t$, hidden state activation $\mathbf{h}_t$ with $\odot$ representing the Hadamard product. Given the spatio-temporal input of the form $\mathbf{u}_t \in \mathbb{R}^{\tau \times m^{\prime}}$, the network then maps the hidden state activation $\mathbf{h}_t$ to a sequence of trading signals $\mathbf{X}_t \in [-1, 1]^{\tau \times N^t}$ by a fully connected layer with time-distributed weights $\mathbf{W}_{\text{dist}}$.

\subsection{Fixed Parameters \& Hyperparameter Optimization}
We perform hyperparameter optimization with 100 iterations of random search for all machine learning models, using the search range as shown in Table \ref{table:hyperparameter_search}. We initiate early stopping with a patience of 25 epochs for the validation loss. We include regularization via dropout \cite{srivastava2014dropout} for the reference DMN, as well as the MLP, CNN and LSTM STMOM models. 
\pagebreak
\begin{table}[H]
\centering
\caption{Fixed Parameters for Machine Learning Models}
\label{table:fixed_parameters}

\begin{tabular}{lccccc}
\toprule
\textbf{Parameters} & \textbf{DMN} & \textbf{SLP} & \textbf{MLP} & \textbf{CNN} & \textbf{LSTM}  \\ 
\midrule
Epochs                 & 100 & 500 & 500 & 500 & 500                                                \\
Patience  & 25 & 25 & 25 & 25 & 25                                                \\
Random Search Iterations & 100 & 100 & 100 & 100 & 100                                                \\
Temporal History & 63 & 5 & 5 & 63 & 63                                                \\          
\hline
\end{tabular}
\end{table}

\begin{table}[H]
\centering
\caption{Hyperparameter Search Range for Machine Learning Models}
\label{table:hyperparameter_search}
\begin{tabular}{lll}
\toprule
\textbf{Hyperparameters}           & \textbf{Random Search Grid}              \\ \midrule
Minibatch Size                     & 32, 64, 128, 256             \\
Dropout Rate                       & 0.1, 0.2, 0.3, 0.4, 0.5           \\
Hidden Layer Size                  & 5, 10, 20, 40, 80, 160       \\
Learning Rate                      & $10^{-5},~ 10^{-4},~ 10^{-3},~ 10^{-2},~ 10^{-1},~ 10^{0}$                       \\
Max Gradient Norm                  & $10^{-4},~ 10^{-3},~ 10^{-2},~ 10^{-1},~ 10^{0},~ 10^{1}$                       \\
L1 Regularisation Weight ($\alpha$) & $10^{-5},~ 10^{-4},~ 10^{-3},~ 10^{-2},~ 10^{-1}, 10^{0}$  \\ \bottomrule
\end{tabular}
\end{table}

\section{Dataset Details}
\label{appendix_b}
All individual instruments in our datasets have less than 10\% missing data. To reduce the effect of extreme outliers, we winsorize all data to limit values to be within 5 times its 252-day exponentially weighted moving standard deviation from its exponentially weighted moving average. \\

\textbf{Equity Index Futures} consists of 12 ratio-adjusted continuous equity index futures contracts obtained from the Pinnacle Data Corp CLC Database. We perform backtesting from 2003 to 2020.
\begin{table}[H]
\centering
\begin{tabular}{cc}
\toprule
\multicolumn{1}{l}{\textbf{Ticker Symbol}} & \textbf{Ticker Description} \\ \midrule
SP                                        & S \& P 500, day session                                       \\
YM                                        & Mini Dow Jones (\$5.00)                                 \\
EN                                       & NASDAQ, MINI                           \\
ER                                       & RUSSELL 2000, MINI                            \\
MD                                        & S\&P 400 (Mini electronic)                             \\
XU                                        & DOW JONES EUROSTOXX50                       \\
XX                                       & DOW JONES STOXX 50                        \\
CA                                       & CAC40 INDEX                      \\
LX                                        & FTSE 100 INDEX                         \\
AX                                        & GERMAN DAX INDEX                             \\
HS                                        & HANG SENG                       \\
NK                                        & NIKKEI INDEX                          \\
\bottomrule
\end{tabular}
\end{table}

\pagebreak
\textbf{US Equities} consists of actively-traded US equities with data obtained from the Center for Research in Security Prices (CRSP). We screen for common stocks of domestic US companies listed on the NYSE, AMEX and NASDAQ with market capitalization from Small (300M-2B) to Mega (>200B) using the Stock Screener provided by Nasdaq. We perform a random sample of 46 stocks from the Financials sector. We perform backtesting from 1990 to 2022. \\\\

\begin{table}[H]
\centering
\begin{tabular}{ccc}
\toprule
\multicolumn{1}{l}{\textbf{Ticker Symbol}} & \textbf{Ticker Description} \\ \midrule
AFG & AMERICAN FINANCIAL GROUP INC NEW  \\
AFL & AFLAC INC  \\
AJG & GALLAGHER ARTHUR J \& CO  \\
BAC & BANK OF AMERICA CORP  \\
C & CITIGROUP INC  \\
CADE & CADENCE BANK  \\
CBSH & COMMERCE BANCSHARES INC  \\
CFR & CULLEN FROST BANKERS INC  \\
CHCO & CITY HOLDING CO  \\
CPF & CENTRAL PACIFIC FINANCIAL CORP  \\
CVBF & C V B FINANCIAL CORP  \\
FITB & FIFTH THIRD BANCORP  \\
GL & GLOBE LIFE INC  \\
IBCP & INDEPENDENT BANK CORP MICH  \\
INDB & INDEPENDENT BANK CORP MA  \\
JEF & JEFFERIES FINANCIAL GROUP INC  \\
JPM & JPMORGAN CHASE \& CO  \\
KEY & KEYCORP NEW  \\
MCO & MOODYS CORP  \\
MCY & MERCURY GENERAL CORP NEW  \\
MKL & MARKEL CORP  \\
NTRS & NORTHERN TRUST CORP  \\
NWLI & NATIONAL WESTERN LIFE GROUP INC  \\
PNC & P N C FINANCIAL SERVICES GRP INC  \\
SBCF & SEACOAST BANKING CORP FLA  \\
SCHW & SCHWAB CHARLES CORP NEW  \\
SEIC & S E I INVESTMENTS COMPANY  \\
SIGI & SELECTIVE INSURANCE GROUP INC  \\
SIVB & S V B FINANCIAL GROUP  \\
SNV & SYNOVUS FINANCIAL CORP  \\
SRCE & 1ST SOURCE CORP  \\
STT & STATE STREET CORP  \\
TFC & TRUIST FINANCIAL CORP  \\
TMP & TOMPKINS FINANCIAL CORP  \\
TRC & TEJON RANCH CO  \\
TROW & T ROWE PRICE GROUP INC  \\
TRST & TRUSTCO BANK CORP NY  \\
UBSI & UNITED BANKSHARES INC  \\
UFCS & UNITED FIRE GROUP INC  \\
USB & U S BANCORP DEL  \\
VALU & VALUE LINE INC  \\
VLY & VALLEY NATIONAL BANCORP  \\
WABC & WESTAMERICA BANCORPORATION  \\
WFC & WELLS FARGO \& CO NEW  \\
WRB & BERKLEY W R CORP  \\
ZION & ZIONS BANCORPORATION N A  \\
\bottomrule
\end{tabular}
\end{table}

\end{document}